\documentclass[12pt]{article}

\usepackage{dcolumn}
\usepackage{bm}
\usepackage{amsfonts,amsmath,amssymb,epsfig}

\begin{document}

\title{Finite amplitude elastic waves \\propagating in compressible solids}
\author{Michel Destrade \& Giuseppe Saccomandi}

\date{2005}
\maketitle

\begin{abstract}
The paper studies the interaction of a
longitudinal wave with transverse waves in general isotropic and
unconstrained hyperelastic materials, including the possibility of
dissipation. The dissipative term chosen is similar to the
classical stress tensor describing a Stokesian fluid and is
commonly used in nonlinear acoustics. The aim of this research is
to derive the corresponding general equations of motion, valid for
any possible form of the strain energy function and to investigate
the possibility of obtaining some general and exact solutions to
these equations by reducing them to a set of ordinary differential
equations. Then the reductions can lead to some exact closed-form
solutions for special classes of materials (here the examples of
the Hadamard, Blatz-Ko, and power-law strain energy densities are
considered, as well as fourth-order elasticity). The solutions
derived are in a time/space separable form and may be interpreted
as generalized oscillatory shearing motions and generalized
sinusoidal standing waves. By means of standard methods of
dynamical systems theory, some peculiar properties of waves
propagating in compressible materials are uncovered, such as for
example, the emergence of destabilizing effects. These latter
features exist for highly nonlinear strain energy functions such
as the relatively simple power-law strain energy, but they cannot
exist in the framework of fourth-order elasticity.
\end{abstract}

\maketitle

\newpage


\section{ Introduction}


Nonlinear elastic wave propagation is a subject of considerable
interest for many natural and industrial applications such as
seismology \cite{sinha}, soft tissue acoustics \cite{Park}, or the
dynamics of elastomers \cite{Kramer}. A detailed study of the
theoretical issues associated with wave phenomena forms the basis
of our understanding of important non-destructive and non-invasive
techniques of investigation such as for instance, the technique of
transient elastography for the analysis of soft solids \cite{CGF}.
The mathematics of wave phenomena is still an active subject of
research where many outstanding problems are waiting for a
definitive systematic treatment. Many papers are devoted to the
study of nonlinear elastic wave propagation and a recent summary
on the status of contemporary research on the subject can be found
in the authoritative review by Norris \cite{Norris}, while an
account of the mathematics of hyperbolic conservation laws can be
found in a book by Dafermos \cite{Dafe}.

It is important to note that in most studies of nonlinear acoustics, the
investigation is usually restricted to \emph{weakly nonlinear} waves.
Indeed, for materials such as metals, rocks, or ceramics, the ratio of the
dynamic displacement to the wavelength is a small parameter and so, the
theories of third-order elasticity or of fourth-order elasticity seem to be
sufficient to account for the nonlinear effects observed in experimental
tests. The situation is however completely different when we consider
elastomeric materials such as those used for vibration isolators or
automobile tires, and when we consider biological materials such as arterial
walls or glands, under physiological or pathological conditions. The
deformation rates occurring in these materials are so large that their
behavior will differ significantly not only from the behavior predicted by
the linear theory of elasticity, but also from the behavior predicted by the
weakly nonlinear theories. It should therefore be important to be able to
derive and to investigate the general equations governing the propagation of
finite amplitude longitudinal and transverse waves in the context of full
nonlinear elasticity, which recovers the weakly nonlinear case as a special
case. Moreover, recent researches in the constitutive behavior of
rubber-like material indicate that the stiffening effect, a peculiar but
real phenomenon occurring at very large strains, cannot be described
accurately when the full polynomial strain energy function is approximated;
also, high order polynomial theories may give rise to numerical difficulties
in the fitting of the material parameters with the experimental data and to
artificial instability phenomena (see Pucci and Saccomandi \cite{PS1} and
Ogden et al. \cite{OgSS04}).

These considerations have led to the present paper which is devoted to the
study of the interaction of a longitudinal wave with transverse waves in
\emph{general isotropic} and \emph{unconstrained hyperelastic} materials and
this also when \emph{dissipation} is taken into account. The aim of our research is
first, to derive the corresponding general equations of motion, valid for
any possible form of the strain energy function; second, to investigate the
possibility of obtaining some general and exact solutions to these equations
by reducing the specific field equations to a set of ordinary differential
equations; and third, to provide, using these reductions, some exact
closed-form solutions for special classes of materials.

To the best of our knowledge, previous studies taking systematically into
account the full nonlinear equations of motion have generally been
restricted to the formal theory of singular surfaces and of acceleration
waves. These studies stem from the fundamental researches initiated by
Hadamard \cite{Hada} in 1903 and were advanced mainly by Ericksen, by
Thomas, and by Truesdell in the 1950s and 1960s (see for example Truesdell
\cite{Tru} and the review by Chen \cite{Chen}). They show that the
conjunction of the nonlinearity in the material response and of the
hyperbolic nature of the governing equations (in the purely elastic case)
leads inevitably to shock formation. The results obtained are general and
exact but they are formal results: in effect, no solution is found
explicitly but rather, conditions are given that have to be fulfilled by
solutions, if they exist.

In contrast, the present article focuses on the explicit determination of
solutions. Our investigation is directly related to the celebrated finite
amplitude elastic motions discovered by Carroll \cite{C1,C1a,C2,C2b,C3,C4,C5}%
. His exact solutions are versatile in their fields of application because
they are valid not only for nonlinearly elastic solids, but also for general
viscoelastic solids \cite{C4}, Reiner-Rivlin fluids \cite{C4, C5}, Stokesian
fluids \cite{C4}, Rivlin-Ericksen fluids \cite{C5}, liquid crystals \cite{C6}%
, dielectrics \cite{C7}, magnetic materials \cite{C8}, etc. They also come
in a great variety of forms, as circularly-polarized harmonic progressive
waves, as motions with sinusoidal time dependence, as motions with
sinusoidal space dependence, etc. Recently the present authors \cite{DeSa05}
extended Carroll's solutions to the case of an incompressible hyperelastic
body in rotation, by considering a new synthetic and very effective method
based on complex variables. In doing so, they discovered a striking analogy
between the equations of motion obtained for a motion general enough to
include all of the above motions, and the equations obtained for the motion
of a nonlinear string, as considered by Rosenau and Rubin \cite{RR}. The
method used by Rosenau and Rubin shows in a simple and direct way that all
(and more) of the different results obtained by Carroll are a direct
consequence of material isotropy and of the Galilean invariance of the field
equations. This is indeed an explanation for their ubiquity and versatility.

The paper is organized in the following manner. In Section II we lay out the
general field equations of nonlinear elasticity. Because dissipation cannot
be neglected for most problems in the dynamics of elastomeric materials and
of soft tissues, we add a simple inelastic tensor of differential type to
the hyperelastic Cauchy stress tensor. This additional term is similar to
the classical stress tensor describing a Stokesian fluid; it is the
dissipative term introduced by Landau and Lifshitz in their monograph on
elasticity theory \cite{LaLi86} and it is commonly used in nonlinear
acoustics (see Norris \cite{Norris}). We point out that this term is
different from the one used in continuum mechanics as a first approximation
of dissipative effects or as a regularization of the hyperbolic equations of
nonlinear elastodynamics for example in numerical computations \cite{Carsten}. 
Then we specialize the equations of motion to the case of one longitudinal
wave and two transverse waves. In Section III we seek solutions in time/space
separable form and derive nonlinear systems of ordinary differential
equations for generalized oscillatory shearing motions and for generalized
sinusoidal standing waves. In Section IV, different classes of materials are
considered (Hadamard, Blatz-Ko, power-law, fourth-order elasticity) and some
exact solutions are provided. Concluding remarks are made in Section V,
which recaps the main results.
In particular, it is emphasized there that a weak 
 nonlinear elasticity theory  (up to the fourth order)
cannot account for some wrinkling phenomenon found in the fully 
nonlinear theory (power-law strain energy).


\section{Preliminaries}



\subsection{Equations of motion}


Consider a hyperelastic body with strain energy density $\Sigma $. Let the
initial and current coordinates of a point of the body, referred to the same
fixed rectangular Cartesian system of axes, be denoted by $\mathbf{X}$ and $%
\mathbf{x}$, respectively. Hence a motion of the body is defined by
\begin{equation}
\mathbf{x}=\mathbf{x}(\mathbf{X},t).  \label{x}
\end{equation}
The response of a homogeneous compressible isotropic elastic solid to
deformations from an undistorted reference configuration is described by the
constitutive relation \cite{Beat87},
\begin{equation}
\mathbf{T}^{E}=2(\frac{I_{2}}{\sqrt{I_{3}}}\frac{\partial \Sigma }{\partial
I_{2}}+\sqrt{I_{3}}\frac{\partial \Sigma }{\partial I_{3}})\mathbf{1}+\frac{2%
}{\sqrt{I_{3}}}\frac{\partial \Sigma }{\partial I_{1}}\mathbf{B}-2\sqrt{I_{3}%
}\frac{\partial \Sigma }{\partial I_{2}}\mathbf{B}^{-1},  \label{T}
\end{equation}
where $\mathbf{T}^{E}$ is the (elastic) Cauchy stress tensor, $\mathbf{1}$
is the unit tensor, $\mathbf{B}$ is the left Cauchy-Green strain tensor
defined by
\begin{equation}
\mathbf{B}:=\mathbf{FF}^{T},  \label{3}
\end{equation}
$\mathbf{F}:=\partial \mathbf{x}/\partial \mathbf{X}$ being the deformation
gradient tensor, and $I_{1}$, $I_{2}$, $I_{3}$ are the first three
invariants of $\mathbf{B}$,
\begin{equation}
I_{1}:=\text{tr }\mathbf{B},\quad I_{2}:=\textstyle{\frac{1}{2}}[I_{1}^{2}-%
\text{tr}(\mathbf{B}^{2})],\quad I_{3}:=\det \mathbf{B}.  \label{invariants}
\end{equation}

To describe the simultaneous effects of thermal and viscoelastic dissipation,
we introduce the following viscous-like stress tensor \cite{LaLi86},
\begin{equation}
\mathbf{T}^{D}=2\eta \lbrack \dot{\mathbf{E}}-\textstyle{\frac{1}{3}}(\text{%
tr }\dot{\mathbf{E}})\mathbf{1}]+(\zeta +\chi )(\text{tr }\dot{\mathbf{E}})%
\mathbf{1},  \label{D}
\end{equation}
where $\mathbf{E}$ is the Green-Lagrange strain tensor,
\begin{equation}
\mathbf{E}:=\textstyle{\frac{1}{2}}(\mathbf{F}^{T}\mathbf{F}-\mathbf{1}),
\label{E}
\end{equation}
and the dot indicates the time derivative. In \eqref{D}, the constant $\eta $
($>0$) is the shear viscosity coefficient, the constant $\zeta $ ($>0$) is
the bulk viscosity coefficient, and the constant $\chi $ ($>0$) is the
coefficient related to the thermal properties of the solid (that is, ambient
temperature, thermal expansion coefficient, and specific heat).

Let $\rho $ and $\rho _{0}$ denote the mass densities of the body measured
in the current configuration and in the reference configuration,
respectively. Then the equation of motions, in the absence of body forces,
are: $\text{div }(\mathbf{T}^{E}+\mathbf{T}^{D})=\rho \ddot{\mathbf{x}}$ in
current form, or equivalently,
\begin{align}
& \text{Div }[\sqrt{I_{3}}(\textbf{T}^{E}+\textbf{T}^{D})(\textbf{F}^{-1})^{T}%
] = \rho _{0}\ddot{\mathbf{x}},  \label{motion} \\
& \dfrac{\partial }{\partial X_{j}}[%
\sqrt{I_{3}}(T_{ik}^{E}+T_{ik}^{D})F_{jk}^{-1}] = \rho
_{0}\dfrac{\partial ^{2}x_{i}}{\partial t^{2}}, \notag
\end{align}
in referential form.


\subsubsection*{Remark on the dissipative stress tensor}


For \textit{incompressible} solids, $\text{det }\mathbf{F} = 1$ at all times
(so that $I_3 = 1$ at all times) and an arbitrary spherical pressure term,
to be determined from initial and boundary conditions, is introduced so that
\begin{equation}  \label{TwithE}
\mathbf{T}^E + \mathbf{T}^D = -p \mathbf{1} + 2 \dfrac{\partial \Sigma}{%
\partial I_1}\mathbf{B} - 2 \dfrac{\partial \Sigma}{\partial I_2} \mathbf{B}%
^{-1} + 2 \eta \dot{\mathbf{E}},
\end{equation}
where $p = p(\mathbf{x},t)$ is a Lagrange multiplier. We note that in order
to study the propagation of finite amplitude motions in viscoelastic \textit{%
incompressible} materials, some authors \cite{HaSa00, HaSa02, DeSa04, HaSa04}
chose to add to the elastic Cauchy stress tensor a viscous term linear in
the stretching tensor $\mathbf{D}$, which is the symmetric part of the
velocity gradient tensor $\dot{\mathbf{F}}\mathbf{F}^{-1}$. In other words,
they chose to write
\begin{equation}  \label{TwithD}
\mathbf{T}^E + \mathbf{T}^D = - p_* \mathbf{1} + 2 \dfrac{\partial \Sigma}{%
\partial I_1}\mathbf{B} - 2 \dfrac{\partial \Sigma}{\partial I_2} \mathbf{B}%
^{-1} + 2\nu \mathbf{D},
\end{equation}
for some Lagrange multiplier $p_*$ and for some constant $\nu$.
However it must be recalled (e.g. Chadwick \cite[p.146]{Chad99})
that $\dot{\mathbf{E}} = \mathbf{F}^{T}\mathbf{DF} \ne \mathbf{D}$
in general. Thus we remark, for the time being, that the
assumptions \eqref{TwithE} and \eqref{TwithD} are not equivalent a
priori.
We discuss these issues further at the end of the next Subsection.


\subsection{Finite amplitude longitudinal and transverse plane waves}


Now we consider the following class of motions,
\begin{equation}
x=u(X,t),\quad y=Y+v(X,t),\quad z=Z+w(X,t),  \label{m1}
\end{equation}
which describes the superposition of a transverse wave, polarized in the ($%
YZ $) plane and propagating in the $X$-direction with a longitudinal wave
propagating in the $X$-direction. Then,
\begin{equation}
\left[ \mathbf{F}\right] _{ij}=
\begin{bmatrix}
u_{X} & 0 & 0 \\
v_{X} & 1 & 0 \\
w_{X} & 0 & 1
\end{bmatrix}
,\quad \left[ \mathbf{F}^{-1}\right] _{ij}=\dfrac{1}{u_{X}}
\begin{bmatrix}
1 & 0 & 0 \\
-v_{X} & u_{X} & 0 \\
-w_{X} & 0 & u_{X}
\end{bmatrix}
,  \label{F}
\end{equation}
and
\begin{equation}
\left[ \dot{\mathbf{E}}\right] _{ij}=\dfrac{1}{2}
\begin{bmatrix}
(u_{X}^{2}+v_{X}^{2}+w_{X}^{2})_{t} & v_{Xt} & w_{Xt} \\
v_{Xt} & 0 & 0 \\
w_{Xt} & 0 & 0
\end{bmatrix}
.  \label{Edot}
\end{equation}
Here and hereafter, a subscript letter denotes partial differentiation (thus
$u_{X}=\partial u/\partial X$, $v_{tt}=\partial ^{2}v/\partial t^{2}$, etc.)
It follows from the definitions \eqref{invariants} that the first three
invariants of $\mathbf{B}$ are given by
\begin{equation}
I_{1}=2+u_{X}^{2}+v_{X}^{2}+w_{X}^{2},\quad
I_{2}=1+2u_{X}^{2}+v_{X}^{2}+w_{X}^{2},\quad I_{3}=u_{X}^{2},
\label{invariants2}
\end{equation}
and consequently that the strain-energy density $\Sigma =\Sigma
(I_{1},I_{2},I_{3})$ is here a function of $u_{X}^{2}$ and $%
v_{X}^{2}+w_{X}^{2}$ alone,
\begin{equation}
\Sigma =\Sigma (u_{X}^{2},v_{X}^{2}+w_{X}^{2}).
\end{equation}

With the expressions \eqref{F} and \eqref{Edot} substituted into the Cauchy
stress \eqref{T} and into the viscous-like tensor \eqref{D}, the equation of
motions \eqref{motion} reduce to
\begin{align}
& \rho _{0}u_{tt}=[(Q_{1}+Q_{2})u_{X}+\textstyle{\frac{1}{2}}(\zeta +\chi +%
\textstyle{\frac{4}{3}}\eta )(u_{X}^{2}+v_{X}^{2}+w_{X}^{2})_{t}]_{X},
\nonumber \\
& \rho _{0}v_{tt}=[Q_{1}v_{X}+\eta v_{Xt}]_{X},  \notag \\
& \rho _{0}w_{tt}=[Q_{1}w_{X}+\eta w_{Xt}]_{X}, \label{equazio}
\end{align}
where the functions $Q_{1}=Q_{1}(u_{X}^{2},v_{X}^{2}+w_{X}^{2})$ and $%
Q_{2}=Q_{2}(u_{X}^{2},v_{X}^{2}+w_{X}^{2})$ are defined by
\begin{equation}
Q_{1}:=2\left( \dfrac{\partial \Sigma }{\partial I_{1}}+\dfrac{\partial
\Sigma }{\partial I_{2}}\right) ,\quad Q_{2}:=2\left( \dfrac{\partial \Sigma
}{\partial I_{2}}+\dfrac{\partial \Sigma }{\partial I_{3}}\right) .
\label{Q1Q2}
\end{equation}
Eqs.~\eqref{equazio} form a system of two coupled nonlinear partial
differential equations, generalizing the system derived by Carroll in \cite
{C1} for an elastic compressible material.

The mathematical treatment of initial-boundary-value problems for
equations such as \eqref{equazio} is not trivial at all.
There is a considerable literature on the subject
and the most updated reference
is the recent paper by Antman and Seidman \cite{Antsed}, to which we
refer for further information. Here we use a semi-inverse
method and we shall consider the possibility to use our
explicit results to solve initial-boundary-value problems only a
posteriori. Obviously, the possibility of general uniqueness
theorems (such as those established in \cite{Antsed})
is quite valuable to give a clear and definitive status
of the solutions found by a semi-inverse method.

Now we differentiate \eqref{equazio} with respect to $X$, and we recast
the resulting equations in the form
\begin{align}
& \rho _{0}U_{tt}=[(Q_{1}+Q_{2})U]_{XX} \notag \\
& \quad \quad +\textstyle{\frac{1}{2}}(\zeta +\chi +%
\textstyle{\frac{4}{3}}\eta )[U^{2}+V^{2}+W^{2}]_{XXt},  \notag
\label{motion1} \\
& \rho _{0}V_{tt}=[Q_{1}V]_{XX}+\eta V_{XXt},  \notag \\
& \rho _{0}W_{tt}=[Q_{1}W]_{XX}+\eta W_{XXt},
\end{align}
where we have introduced the new functions: $U:=u_{X}$, $V:=v_{X}$, $W:=w_{X}
$.

Further, these equations can be rewritten in an even more compact form, by
introducing the complex function $\Lambda$, with modulus $\Omega$ and
argument $\theta$, defined by
\begin{equation}  \label{Lambda}
\Lambda(X,t) = \Omega(X,t)\text{e}^{i\theta(X,t)}:= V + iW,
\end{equation}
so that
\begin{align}
& V = \Re (\Lambda) = \Omega \cos \theta ,\quad W = \Im (\Lambda)
= \Omega \sin \theta, \notag \\
& V^2 + W^2 = \Omega^2.
\end{align}
Then the three equations \eqref{motion1} are
\begin{align}  \label{motion2}
& \rho_0 U_{tt} = [(Q_1 + Q_2) U]_{XX} +\textstyle{\frac{1}{2}} (\zeta +
\chi + \textstyle{\frac{4}{3}}\eta) [U^2 + \Omega^2]_{XXt},  \notag \\
& \rho_0 \Lambda_{tt} = [Q_1 \Lambda]_{XX} + \eta \Lambda_{XXt}.
\end{align}
Now also, the invariants $I_1$, $I_2$, $I_3$ found in \eqref{invariants2}
are written as
\begin{equation}  \label{invariants3}
I_1 = 2 + U^2 + \Omega^2, \quad I_2 = 1 + 2U^2 + \Omega^2, \quad I_3 = U^2,
\end{equation}
so that $Q_1 = Q_1(U^2, \Omega^2)$ and $Q_2 = Q_2(U^2, \Omega^2)$ or, more
explicitly,
\begin{equation}
Q_1 = 2 \dfrac{\partial \Sigma}{\partial (\Omega^2)}, \quad Q_2 = 2 \left(%
\dfrac{\partial \Sigma}{\partial (U^2)} -\dfrac{\partial \Sigma}{\partial
(\Omega^2)} \right).
\end{equation}

These equations have been developed in the \emph{purely elastic} case of a
\emph{single} shear wave by Carman and Cramer \cite{CaCr92}, who determined
some numerical and asymptotic solutions.


\subsubsection*{Remark on finite amplitude motions in incompressible
materials}


In \textit{incompressible} solids, $\text{det }\mathbf{F}=1$ at
all times and so by \eqref{F}$_{1}$, $u_{X}=1$. Then we find
\begin{equation}
\left[ \dot{\mathbf{E}}\right] _{ij}=\dfrac{1}{2}
\begin{bmatrix}
(v_{X}^{2}+w_{X}^{2})_{t} & v_{Xt} & w_{Xt} \\
v_{Xt} & 0 & 0 \\
w_{Xt} & 0 & 0
\end{bmatrix},  \label{ri1}
\end{equation}
and
\begin{equation}
\left[ \mathbf{D}\right] _{ij}=\dfrac{1}{2}
\begin{bmatrix}
0 & v_{Xt} & w_{Xt} \\
v_{Xt} & 0 & 0 \\
w_{Xt} & 0 & 0
\end{bmatrix}
.  \label{ri1b}
\end{equation}
It then follows that
\begin{equation}
\text{Div }[(-p\textbf{1}+2\eta \dot{\textbf{E}})(\textbf{F}^{-1})^{T}]=-%
\text{Grad }\hat{p}+\eta v_{XXt}\mathbf{j}+\eta w_{XXt}\mathbf{k},
\end{equation}
where $\hat{p}:=p-(v_{X}^{2}+w_{X}^{2})_{t}$, and that
\begin{equation}
\text{Div }[(-p_{\ast }\textbf{1}+2\nu \textbf{D})(\textbf{F}^{-1})^{T}]=-%
\text{Grad }p_{\ast }+\nu v_{XXt}\mathbf{j}+\nu w_{XXt}\mathbf{k}.
\end{equation}
Now we see that the equations of motion \eqref{motion} are the same whether
the choice \eqref{TwithE} or the choice \eqref{TwithD} is made. The two
approaches in modeling the dissipative effects are reconciled, at least as
far as finite amplitude motions in incompressible materials are concerned.



\section{Solutions in separable form}


The general solution of \eqref{motion2} may be found only via numerical
methods. To obtain some simple analytical information here, we look for
reductions of the governing equations \eqref{motion2} to a set of ordinary
differential equations. The underlying idea is to search for $\Lambda (X,t)$%
, defined in \eqref{Lambda}, in a separable form such as
\begin{equation}
\Omega (X,t)=\Omega _{1}(X)\Omega _{2}(t),\quad \theta (X,t)=\theta
_{1}(X)+\theta _{2}(t),  \label{ses3}
\end{equation}
where $\Omega _{1}$, $\theta _{1}$ are functions of space only and $\Omega
_{2}$, $\theta _{2}$ are functions of time only. Some lengthy computations,
not reproduced here, show that in only two general cases can equations
\eqref{motion2} be separated in such way for \textit{any} material response.

The results of this Section generalize the results of Carroll \cite{C1a, C3}
by considering more general types of displacement fields and by including
viscoelastic effects. We also acknowledge that the solutions presented here
are a generalization of the non-linear three-dimensional motions of an
elastic string found by Rosenau and Rubin \cite{RR} (see \cite{DeSa05} for a
different generalization, to finite amplitude waves in rotating
incompressible elastic bodies.)


\subsection{Generalized oscillatory shearing motions}


The first general type of separable solution is of the form,
\begin{equation}
\Lambda (X,t)=[\psi (t)+i\phi (t)]k\text{e}^{i(kX-\theta (t))},
\label{ses13}
\end{equation}
where $k$ is a constant and $\psi$, $\phi$, $\theta$ are arbitrary real
functions of time. Here $\Lambda $ is indeed of the separable form given by %
\eqref{Lambda}, with
\begin{align}
& \Omega_1(X) = k = \text{const}., & & \Omega_2(t) = [\phi(t)^2 +
\psi(t)^2]^{\textstyle{\frac{1}{2}}},  \notag \\
& \theta_1(X) = kX, & & \theta_2(t) = \theta(t) + \tan^{-1}[\phi(t) /
\psi(t)].
\end{align}
Separating real and imaginary parts, we find that the transverse
displacement field $(v,w)$ derived from \eqref{ses13} is
\begin{align}
& v(X,t) =\phi (t)\cos (kX-\theta (t))+\psi (t)\sin (kX-\theta (t)),  \notag
\\
& w(X,t) =\phi (t)\sin (kX-\theta (t))-\psi (t)\cos (kX+\theta (t)).
\label{field1}
\end{align}
In this case,
\begin{equation}
\Omega^2 = k^2 [\phi^2 (t) + \psi^2(t)],
\end{equation}
is a function of time only.

Now the first equation of motion \eqref{motion2}$_{1}$ reduces to
\begin{equation}
\rho _{0}U_{tt}=[(Q_{1}+Q_{2})U]_{XX}+\textstyle{\frac{1}{2}}(\zeta +\chi +%
\textstyle{\frac{4}{3}}\eta )[U^{2}]_{XXt},
\end{equation}
where
\begin{equation}
Q_{1}=Q_{1}(U^{2}(X,t),\Omega ^{2}(t)),\quad Q_{2}=Q_{2}(U^{2}(X,t),\Omega
^{2}(t)).  \label{Q_1Q_2}
\end{equation}
On the other hand, the second equation of motion \eqref{motion2}$_{2}$
reduces to
\begin{equation}
\rho _{0}\Lambda _{tt}=Q_{1}\Lambda _{XX}+\eta \Lambda _{XXt}.  \label{ses16}
\end{equation}
Using the separable form \eqref{ses13} for $\Lambda $, factoring out the
exponential term, and separating the real and imaginary parts yields
\begin{align}
& \rho _{0}(\phi \theta ^{\prime \prime }-\psi \theta ^{^{\prime }2}+2\phi
^{\prime }\theta ^{\prime }+\psi ^{\prime \prime })+k^{2}Q_{1}\psi
+k^{2}\eta (\phi \theta ^{\prime }+\psi ^{\prime })=0,  \notag  \label{ses17}
\\
& \rho _{0}(\phi ^{\prime \prime }-2\psi ^{\prime }\theta ^{\prime }-\phi
\theta ^{^{\prime }2}-\psi \theta ^{\prime \prime })+k^{2}Q_{1}\phi
+k^{2}\eta (\phi ^{\prime }-\psi \theta ^{\prime })=0.
\end{align}
This is a nonlinear system of two \textit{ordinary} differential equations
in the three unknowns $\phi (t)$, $\theta (t)$, and $\psi (t)$ either: when $%
Q_1$ is a constant independent of its arguments $U^2(X,t)$ and $\Omega^2(t)$%
, or: when $Q_1 = Q_1(t)$. By \eqref{Q1Q2}$_{1}$, this latter condition is
satisfied if and only if $U(X,t)=U(t)$, a function of time only. It then
follows by \eqref{Q_1Q_2}$_{2}$ that $Q_{2}=Q_{2}(t)$ also. By substitution
into the first equation of motion \eqref{motion2}$_{1}$, we obtain $\rho
_{0}U^{\prime \prime }=0$. Then the longitudinal displacement field is given
by
\begin{equation}
U(t)=C_{1}t+C_{2}=u_{X}, \label{ses15}
\end{equation}
so that
\begin{equation}
u(X,t)=(C_{1}t+C_{2})X+C_{3},  \label{ses15b}
\end{equation}
where $C_{1}$, $C_{2}$, $C_{3}$ are constants. This is a
\emph{homogeneous accelerationless} motion.

Hence we showed that for any compressible elastic material with viscoelastic
dissipative part defined as in \eqref{D}, maintained in a state of
longitudinal simple extension, the special transverse waves \eqref{field1}
may always propagate.

Because \eqref{ses17} is a system of only two equations in three unknowns,
there is a certain freedom in considering special classes of solutions. One
such class of special solutions is obtained by considering $\psi :=0$ in %
\eqref{field1}, leading to
\begin{align}
& v(X,t)=\phi (t)\cos (kX-\theta (t)),  \notag \\
& w(X,t)=\phi (t)\sin (kX-\theta (t)),  \label{string}
\end{align}
a direct generalization of the classical damped harmonic
circularly-polarized wave solution,
\begin{align}
& v(X,t)=A\text{e}^{-ht}\cos (kX-\omega t),  \notag \\
& w(X,t)=A\text{e}^{-ht}\sin (kX-\omega t),  \label{damp}
\end{align}
where $A$, $h$, and $\omega $ are suitable constants. For these
special motions at $\psi :=0$, the system \eqref{ses17} reduces to
\begin{align}
& \rho_0 (\phi \theta^{\prime\prime}+ 2\phi^{\prime}\theta^{\prime}) + \eta
k^2 \phi \theta^{\prime}= 0,  \notag \\
& \rho_0 (\phi^{\prime\prime}- \phi \theta^{\prime}{}^{2}) + k^2 Q_1 \phi +
\eta k^2 \phi^{\prime}= 0.  \label{ses20}
\end{align}
The equation \eqref{ses20}$_{1}$ admits the first integral
\begin{equation}
\phi^2 \theta^{\prime}= E \text{e}^{-\frac{\eta k^2}{\rho_0} t},
\label{firstIntegral}
\end{equation}
where $E$ is a constant of integration.

If $E =0$, then $\theta(t) = \theta_0$, a constant. We end up with standing
waves,
\begin{equation}
v(X,t) = \phi(t) \cos(kX - \theta_0), \quad w(X,t) = \phi(t) \sin(kX -
\theta_0),  \label{ses23}
\end{equation}
where, according to \eqref{ses20}$_2$, $\phi$ is a solution to
\begin{equation}
\rho_0 \phi^{\prime\prime}+ k^2 \eta \phi^{\prime}+ k^2 Q_1(k^2 \phi^2, t)
\phi = 0,  \label{ses24}
\end{equation}
which is the equation of a damped nonlinear oscillator.

If $E \ne 0$, then solving \eqref{firstIntegral} for $\theta^{\prime}$, we
end up with the single equation \eqref{ses20}$_2$, which reduces to
\begin{equation}
\rho_0 \phi^{\prime\prime}+ k^2 \eta \phi^{\prime}+ k^2 Q_1 \phi - \dfrac{%
\rho_0 E^2}{\phi^3} \text{e}^{-\frac{2\eta k^2}{\rho_0} t} = 0.
\label{ses21b}
\end{equation}
In the elastic case ($\eta =0$), this equation describes the plane motion of
a particle in a central force field. In general ($\eta \ne 0$), it is a
nonlinear and non-autonomous differential equation.

For \textit{harmonic wave propagation}, $\theta (t)$ is taken as $\theta(t)
= \omega t$. By \eqref{firstIntegral}, $\phi$ is proportional to $\text{e}^{-%
\frac{\eta k^2}{2\rho_0} t}$ and describes the usual linear damping
function. However by substitution in \eqref{ses20}$_{2}$, we see that this
situation is possible only if the material response is such that $Q_1$ is a
constant,
\begin{equation}
Q_1 = 2\left( \dfrac{\partial \Sigma }{\partial I_1} + \dfrac{\partial
\Sigma }{\partial I_2}\right) = \text{const.} = \rho_0 \dfrac{\omega^2}{k^2}
+ \dfrac{\eta^2 k^2}{4\rho_0}.  \label{ses22}
\end{equation}

In this Subsection we analyzed the solutions \eqref{string} which are the
analogue of the Class I solutions presented by Rosenau and Rubin \cite{RR1}
for purely elastic strings. Qualitatively the solutions describe particles
moving in a helical path which lies on a cylindrical surface of time-varying
radius.


\subsection{Generalized sinusoidal standing waves}


We now consider the second class of solutions in separable form, namely
\begin{equation}
\Lambda (X,t)=[(i\phi (X)+\psi (X))\theta ^{\prime }(X)+(\phi ^{\prime
}(X)-i\psi ^{\prime }(X))]\text{e}^{i(\omega t+\theta (X))},  \label{ses14}
\end{equation}
where $k$ is a constant and $\psi (X)$, $\phi (X)$, $\theta (X)$ are
arbitrary functions of space. Here $\Lambda (X,t)$ is indeed of the
separable form \eqref{Lambda}, with
\begin{align}
& \Omega_1(X) = [(\phi^{\prime}+ \psi \theta^{\prime})^2 + (\phi
\theta^{\prime}- \psi^{\prime})^2]^{\textstyle{\frac{1}{2}}},
\quad \Omega_2(t) = 1 = \text{const}., \notag \\
& \theta_1(X) = \theta + \tan^{-1}[(\phi \theta^{\prime}-
\psi^{\prime})/(\psi \theta^{\prime}+ \phi^{\prime})], \quad
\theta_2(t)= \omega t.
\end{align}
The transverse displacement field ($v,w$) corresponding to \eqref{ses14} is
\begin{align}
& v(X,t) = \phi(X) \cos (\omega t + \theta(X)) + \psi(X) \sin(\omega t +
\theta(X)),  \notag \\
& w(X,t) = \phi(X) \sin (\omega t+\theta(X)) - \psi(X) \cos(\omega t +
\theta(X)).  \label{ses14b}
\end{align}

In this case, $\Omega = \Omega_1(X)$, a function of $X$ only, and the first
equation of motion \eqref{motion2}$_1$ reduces to
\begin{equation}
\rho _{0}U_{tt}=[(Q_{1}+Q_{2})U]_{XX}+\textstyle{\frac{1}{2}}(\zeta +\chi +%
\textstyle{\frac{4}{3}}\eta )[U^{2}]_{XXt},  \label{eqred1}
\end{equation}
where
\begin{equation}
Q_1 = Q_1(U^2, \Omega_1^2(X)), \quad Q_2 = Q_2(U^2, \Omega_1^2(X)).
\end{equation}

Now, introducing \eqref{ses14} into the second equation of motion %
\eqref{motion2}$_2$ and separating the real part from the imaginary part, we
obtain
\begin{multline}  \label{bigEq1}
- \rho_0 \omega^2 (\psi \theta^{\prime}+ \phi^{\prime}) = (\psi
\theta^{\prime}+ \phi^{\prime}) (Q_1)_{XX} + 2
(2\psi^{\prime}\theta^{\prime}+ \psi \theta^{\prime\prime}+
\phi^{\prime\prime}- \phi \theta^{^{\prime}2}) (Q_1)_X \\
+ (3\psi^{\prime\prime}\theta^{\prime}+ 3\psi^{\prime}\theta^{\prime\prime}+
\psi \theta^{\prime\prime\prime}- \psi \theta^{^{\prime}3} -
3\phi^{\prime}\theta^{^{\prime}2} - 3 \phi
\theta^{\prime}\theta^{\prime\prime}+ \phi^{\prime\prime\prime})Q_1 \\
- \eta \omega (3 \phi^{\prime\prime}\theta^{\prime}+ 3
\phi^{\prime}\theta^{\prime\prime}+ \phi \theta^{\prime\prime\prime}- \phi
\theta^{^{\prime}3} + 3 \psi^{\prime}\theta^{^{\prime}2} + 3 \psi
\theta^{\prime}\theta^{\prime\prime}- \psi^{\prime\prime\prime}),
\end{multline}
and
\begin{multline}  \label{bigEq2}
- \rho_0 \omega^2 (\phi \theta^{\prime}- \psi^{\prime}) = (\phi
\theta^{\prime}- \psi^{\prime}) (Q_1)_{XX} + 2
(2\phi^{\prime}\theta^{\prime}+ \phi \theta^{\prime\prime}-
\psi^{\prime\prime}+ \psi \theta^{^{\prime}2}) (Q_1)_X \\
+ (3\phi^{\prime\prime}\theta^{\prime}+ 3\phi^{\prime}\theta^{\prime\prime}+
\phi \theta^{\prime\prime\prime}- \phi \theta^{^{\prime}3} +
3\psi^{\prime}\theta^{^{\prime}2} + 3 \psi
\theta^{\prime}\theta^{\prime\prime}- \psi^{\prime\prime\prime})Q_1 \\
+ \eta \omega (3 \psi^{\prime\prime}\theta^{\prime}+ 3
\psi^{\prime}\theta^{\prime\prime}+ \psi \theta^{\prime\prime\prime}- \psi
\theta^{^{\prime}3} - 3 \phi^{\prime}\theta^{^{\prime}2} - 3 \phi
\theta^{\prime}\theta^{\prime\prime}+ \phi^{\prime\prime\prime}).
\end{multline}
A sufficient condition for the equations \eqref{bigEq1} and \eqref{bigEq2}
to compose a nonlinear system of two \textit{ordinary} differential
equations in the three unknowns $\phi$ , $\psi$, and $\theta$ is that the
longitudinal field depends only on $X$: $U(X,t)\equiv U(X)$. In this case, $%
Q_1 = Q_1(X)$, $Q_2 = Q_2(X)$ and \eqref{eqred1} reduces to
\begin{equation}
\lbrack (Q_{1}(X)+Q_{2}(X))U(X)]^{\prime \prime }=0.  \label{cm0b}
\end{equation}

Further progress is made by fixing one of the three unknown functions, $%
\theta :=0$, say. Then the governing equations \eqref{bigEq1} and %
\eqref{bigEq2} reduce to
\begin{equation}
(Q_{1}\phi ^{\prime })^{\prime }+\rho _{0}\omega ^{2}\phi =-\eta \omega \psi
^{\prime \prime },\quad (Q_{1}\psi ^{\prime })^{\prime }+\rho _{0}\omega
^{2}\psi =\eta \omega \phi ^{\prime \prime }.  \label{cm0}
\end{equation}
At $\eta =0$, these equations are formally equivalent to the equations
derived by Carroll \cite{C2}, with the difference that the materials
response functions $Q_{1}$ and $Q_{2}$ depend not only on $\Omega ^{2}$ (the
amount of shear) but also on $U$. In Carroll \cite{C5} (see system 5.12 in
that reference), the attention is restricted to \emph{incompressible} fluids
and solids, for which $U(X)$ must be constant.

The solutions considered in this sub-Section contain the Class II solutions
of Rosenau and Rubin \cite{RR1}.


\section{Specific materials}


Now that we have established the possibility of reducing the general
nonlinear partial differential equations to ordinary differential equations,
we must specify constitutive relations in order to make progress.

\subsection{Hadamard materials}


Finite amplitude motions in \textit{elastic} Hadamard materials have been
thoroughly studied, see John \cite{John66} and Boulanger et al. \cite{BoHT94}. 
The analysis conducted in this paper allows for a treatment of a
\emph{dissipative} Hadamard solid.

The Hadamard strain energy function is defined by
\begin{equation}
2\Sigma = C(I_1-3) + D(I_2 - 3) + G(I_3),  \label{h1}
\end{equation}
where $G(I_3)$ is a material function and $C$, $D$ are material constants
such that \cite{BoHT94} $C>0$, $D \ge 0$, or $C \ge 0$, $D>0$. The
connection with the Lam\'e constants $\lambda$ and $\mu$ of the linear
theory of isotropic elasticity is made through the relations:
\begin{equation}
C = 2 \mu + G^{\prime}(1), \quad D = - \mu - G^{\prime}(1), \quad
2G^{\prime\prime}(1) = \lambda + 2\mu.  \label{h2}
\end{equation}

For this material, the functions $Q_1$ and $Q_2$ defined in \eqref{Q1Q2} are
\begin{equation}
Q_1 = C + D, \quad Q_2 = D + G^{\prime}(U^2),  \label{h3}
\end{equation}
and the equations of motion \eqref{motion2} reduce to
\begin{align}
& \rho_0 U_{tt} = \{ [C + 2D + G^{\prime}(U^2)] U \}_{XX} \notag \\
& \quad \quad \quad + {\textstyle\frac{1}{2}}(\zeta + \chi +
\textstyle{\frac{4}{3}} \eta)(U^2 + \Omega^2)_{tXX},
\notag \\
& \rho_0 \Lambda_{tt} = (C + D) \Lambda_{XX} + \eta \Lambda_{tXX}.
\label{h4}
\end{align}

We point out that the equation governing the propagation of the transverse
waves \eqref{h4}$_2$ is always linear, for all Hadamard materials, elastic
\emph{or} dissipative. Hence we see at once that classical harmonic (damped and
attenuated) transverse waves are always possible for these materials.

The material function $G(I_{3})$ accounts for the effects of
compressibility. For example, Levinson and Burgess \cite{LB} proposed the
following explicit form,
\begin{equation}
G(I_{3})=(\lambda +\mu )(I_{3}-1)-2(\lambda +2\mu )(\sqrt{I_{3}}-1).
\label{h5}
\end{equation}
This function leads here to a remarkably simple system of equations:
\begin{align}
& \rho _{0}U_{tt}=(\lambda +2\mu )U_{XX}+{\textstyle\frac{1}{2}}(\zeta +\chi
+\textstyle{\frac{4}{3}}\eta )(U^{2}+\Omega ^{2})_{tXX},  \notag \\
& \rho _{0}\Lambda _{tt}=\mu \Lambda _{XX}+\eta \Lambda _{tXX}.  \label{h6}
\end{align}

Now we take a look at some of the solutions investigated in the previous
Section. Classical \textit{damped harmonic circularly polarized waves} such
as \eqref{damp} are possible in a dissipative Hadamard material because then
$\Omega = kAe^{-ht}$, a function of time only, and $Q_1=C+D=\mu$, a
constant. The corresponding dispersion equation \eqref{ses22} is here
\begin{equation}
\mu = \rho_0 \dfrac{\omega^2}{k^2} + \dfrac{\eta^2 k^2}{4\rho_0}.
\end{equation}
The associated longitudinal motion may be either the homogeneous
accelerationless motion \eqref{ses15} or any solution to the nonlinear
equation \eqref{h4}$_1$, here:
\begin{equation}
\rho_0 U_{tt} = \{[G^{\prime}(U^2) - G^{\prime}(1)]U\}_{XX} + {\textstyle%
\frac{1}{2}}(\zeta + \chi + \textstyle{\frac{4}{3}}\eta) (U^2)_{tXX},
\label{h6a}
\end{equation}
which for the Levinson and Burgess choice \eqref{h5} of $G$ is:
\begin{equation}
\rho_0 U_{tt} = (\lambda + 2\mu) U_{XX} + {\textstyle\frac{1}{2}}(\zeta +
\chi + \textstyle{\frac{4}{3}} \eta)(U^2)_{tXX}.
\end{equation}

Now consider the  \textit{sinusoidal standing waves} of Eq.\eqref{ses14b} 
at $\theta (X)=0$. 
When the longitudinal field $U$ depends on
$X$ only, their behavior is governed by Eqs.\eqref{cm0}, here:
\begin{equation}
\mu \phi ^{\prime \prime }=\rho _{0}\omega ^{2}\phi +\eta \omega \psi
^{\prime \prime },\quad \mu \psi ^{\prime \prime }=\rho _{0}\omega ^{2}\psi
-\eta \omega \phi ^{\prime \prime },  \label{h6b}
\end{equation}
or equivalently, by the single complex equation
\begin{equation}
(\mu -i\eta \omega )(\phi +i\psi )^{\prime \prime }=\rho \omega ^{2}(\phi
+i\psi ).  \label{h6c}
\end{equation}
This equation possesses the following class of attenuated solutions,
\begin{align}
& \phi (X)=e^{-\alpha X}[k_{1}\sin (\beta X)+k_{2}\cos (\beta X)],  \notag \\
& \psi (X)=e^{-\alpha X}[k_{3}\cos (\beta X)+k_{4}\sin (\beta X)],
\label{h6d}
\end{align}
where $k_1$, $k_2$ are disposable constants and
\begin{equation}
\alpha ,\beta =\sqrt{\frac{\rho _{0}\omega ^{2}}{2(\mu ^{2}+\eta ^{2}\omega
^{2})}\left[ \sqrt{\mu ^{2}+\eta ^{2}\omega ^{2}}\pm \mu \right] }.
\label{h6e}
\end{equation}
These solutions may be used to solve some simple boundary value problems.
For example, consider the case of semi-infinite body bounded by a vibrating
rigid plate at $X=0$, and take the velocity components of the plate as
\begin{equation}
\dot{x}=0,\quad \dot{y}=V\omega \cos (\omega t),\quad \dot{z}=V\omega \sin
(\omega t).  \label{h6aa}
\end{equation}
Then $\phi (0)=0$, $\psi (0)=V$, and the associated transverse displacements
are
\begin{align}
& v(X,t)=e^{-\alpha X}[\sin (\beta X)\cos (\omega t)+V\cos (\beta X)\sin
(\omega t)],  \notag \\
& w(X,t)=e^{-\alpha X}[\sin (\beta X)\sin (\omega t)-V\cos (\beta
X)\cos (\omega t)]. \label{bbb}
\end{align}
The associated longitudinal displacement independently satisfies \eqref{cm0b}%
, here:
\begin{equation}
\{[G^{\prime }(U^{2})-G^{\prime }(1)]U\}^{\prime \prime }=0,
\end{equation}
which for the Levinson and Burgess choice \eqref{h5} of $G$ is simply: $%
U^{\prime \prime }=0$. Then the basic displacement \eqref{bbb} may
be superimposed upon an axial static stretch or a homogeneous
motion in the axial direction. In a similar way, it is possible to
solve the case of a slab fixed at $X=0$ and oscillating at $X=L$,
or it is possible to use the various integration constants to fix
the stress traction or shear stress at the boundary of a vibrating
half-space or of a slab.


\subsection{Blatz-Ko materials}


Experiments on compressible polyurethane rubber lead Blatz and Ko \cite
{BlKo62} to propose some strain energy functions which have since received
much attention, see for instance, Beatty \cite{Beat87}. In particular, two
reduced forms of the Blatz-Ko general constitutive equation were deemed
appropriate to model certain polyurethane rubber samples. One is the
Blatz-Ko strain energy function for a solid, polyurethane rubber,
\begin{equation}
\Sigma = \dfrac{\mu}{2} [I_1 -3 + \beta (I_3^{1/\beta} - 1)],
\end{equation}
where $\mu$ and $\beta$ are constants. Direct comparison with \eqref{h1}
shows that this material is a special Hadamard material, with the
identifications: $C = \mu$, $D=0$, $G(I_3) = -\mu \beta I_3^{1/\beta}$. The
other is the \textit{Blatz-Ko strain energy function for a foamed,
polyurethane elastomer},
\begin{equation}
\Sigma = \dfrac{\mu}{2} \left[\dfrac{I_2}{I_3} - 3 + \beta(I_3^{1/\beta} - 1)%
\right],  \label{h7}
\end{equation}
where $\mu$ and $\beta$ are constants. The continuity with linear elasticity
requires that $\mu$ is the infinitesimal shear modulus and that $\beta =
(1-2 \nu)/ \nu$, where $\nu$ is the infinitesimal Poisson ratio. Blatz and
Ko's experiments showed that typically, $\nu =1/4$ (and so, $\beta=2$).

For this latter material, the functions $Q_1$ and $Q_2$ defined in %
\eqref{Q1Q2} are
\begin{equation}
Q_1 = \mu U^{-2}, \quad Q_2 = -\mu U^{-2} - \mu (1 + \Omega^2)U^{-4} + \mu U^%
\frac{2(\beta -1)}{\beta},  \label{h7a}
\end{equation}
and the equations of motion \eqref{motion2} reduce to
\begin{align}
& \rho_0 U_{tt} = - \mu \left[(1 + \Omega^2)U^{-3} - U^\frac{3 \beta -2}{%
\beta}\right]_{XX} \notag \\
& \quad \quad \quad + {\textstyle\frac{1}{2}} (\zeta + \chi + \textstyle{%
\frac{4}{3}}\eta)(U^2 + \Omega^2) _{tXX},  \notag \\
& \rho _{0}\Lambda _{tt} = \mu(\Lambda U^{-2})_{XX} + \eta \Lambda_{tXX}.
\label{h8}
\end{align}
Here, and in contrast with the case of Hadamard materials, the transverse
wave is coupled to the longitudinal wave.

First, consider the \textit{generalized oscillatory shearing motions} of
Section III.A. As seen there, they are governed by nonlinear ordinary
differential equations when the longitudinal displacement $u$ is the
homogeneous accelerationless motion of \eqref{ses15}$_2$. Then the standing
waves of \eqref{ses23} are governed by
\begin{equation}
\rho_0 \phi^{\prime\prime}+ \eta k^2 \phi^{\prime}+ \dfrac{\mu k^2}{(C_1 t +
C_2)^2} \phi = 0.  \label{h10}
\end{equation}

At $C_1 = 0$, this is the classical equation of a damped oscillator, whose
type of motion (exponential and/or sinusoidal) is decided by the sign of the
quantity: $(\eta k C_2)^2 - 4 \rho_0 \mu$; hence the nature of the
transverse motion depends on the longitudinal extension via the parameter $%
C_2$.

At $C_ 1\neq 0$ (linear stretch rate), 
we perform the change of variables $\zeta = C_1 t + C_2$,
so that \eqref{h10} reduces to
\begin{equation}  \label{h12b}
\zeta^2 \dfrac{\text{d}^2 \phi}{\text{d} \zeta^2} 
 + 2 \Upsilon \zeta^2 \dfrac{\text{d}\phi}{\text{d}\zeta} 
  + \dfrac{\Gamma}{4}  \phi = 0.
\end{equation}
where the quantities $\Upsilon$, $\Gamma$ are defined by
\begin{equation}
 \Upsilon := \dfrac{\eta k^2}{2\rho_0 C_1}, 
\quad
 \Gamma :=\frac{4 \mu k^2}{\rho_0 C_1^2}.
\end{equation}
In the purely elastic case ($\eta =0$), 
$\Upsilon$ vanishes and the solution to this equation is
quite simple: we find that for
\begin{align}
& \Gamma < 1, & & \phi(\zeta) = k_1 \zeta^{\lambda_+} + k_2
\zeta^{\lambda_-}, \quad \text{where} \quad \lambda_\pm = (1 \pm \sqrt{1 -
\Gamma})/2;  \notag \\
& \Gamma = 1, & & \phi(\zeta ) = k_1 \sqrt{\zeta} + k_2 \sqrt{\zeta} \ln
\zeta;  \notag \\
& \Gamma > 1, & & \phi(\zeta) = k_1 \sqrt{\zeta} \cos(\sqrt{\Gamma - 1} \ln
\zeta) + k_2 \sqrt{\zeta} \sin(\sqrt{\Gamma - 1} \ln \zeta);  \label{h12c}
\end{align}
where $k_1$ and $k_2$ are integration constants. We note that all these
solutions \emph{blow up} as $t \rightarrow \infty$ (equivalent to $\zeta
\rightarrow \infty$); that for $\Gamma \leq 1$, the solutions are monotonic
increasing; and that for $\Gamma > 1$, the solutions have a slight oscillatory
character. 
In the viscoelastic case ($\eta \neq 0$), equation \eqref{h12b}
is solvable in terms of special functions and a richer 
variety of solutions emerges, because now 
the solutions are \emph{not necessarily unbounded}.
Moreover, although they are damped solutions, they 
do \emph{not necessarily vanish with time}. 
In order to solve  equation \eqref{h12b} explicitly, 
we introduce the function $f$ defined by 
\begin{equation}
	\phi(\zeta) = e^{-\Upsilon \zeta} f(2 \Upsilon \zeta),
\end{equation}
and we find that it satisfies a Whittaker equation:
\begin{equation}
	f''(2 \Upsilon \zeta) 
	 + \left(
	      -\dfrac{1}{4} 
	      + \dfrac{\Gamma}{4(2 \Upsilon \zeta)^2} 
	   \right)
	f(2 \Upsilon \zeta) = 0.
\end{equation}
It follows that the solution to \eqref{h12b}
is expressed in terms of Whittaker's functions as:
\begin{equation}  \label{h12d}
\phi(\zeta) = e^{ -\Upsilon \zeta}  
 \left[
     k_1 W_{0, \frac{1}{2}\sqrt{1 - \Gamma}}   \left(2\Upsilon \zeta\right) 
   + k_2 M_{0, \frac{1}{2}\sqrt{1 - \Gamma}} \left(2\Upsilon \zeta\right) 
 \right],
\end{equation}
where $k_1$ and $k_2$ are integration constants.
To illustrate the influence of the stretch rate in the 
$X$-direction (through the constant $C_1$) 
upon the behavior of the solutions, we 
consider that the material is unstretched at $t=0$
(so that $C_2 =1$ and $\phi(t=0)=0$); 
we take $\phi'(t=0) = 1$, 
$\mu k^2 / \rho_0 = 1$, $\eta = \mu /10$ 
(so that $\Upsilon = 1/(20 C_1)$ and $\Gamma = 4/C_1^2$); 
and we let $C_1 = 0.2$, $1$, $1.5$. 
Figure 1 shows that for a ``small'' rate, the solution undergoes 
a few oscillations before it vanishes completely; as $C_1$ increases, 
the oscillations disappear but the solution tends to an
increasing finite asymptotic value. 
We argue that this behavior can be related to the problem 
of a tensile impact: 
when the material is plucked at ``low'' speed, then a few 
oscillations take place before the material returns to its 
original state;
when the material is plucked at ``high'' speed, then 
the material can accomodate an asymptotic transverse wave.

\begin{figure}[!t]
 \setlength{\unitlength}{1mm}
 \centering 
\epsfig{figure=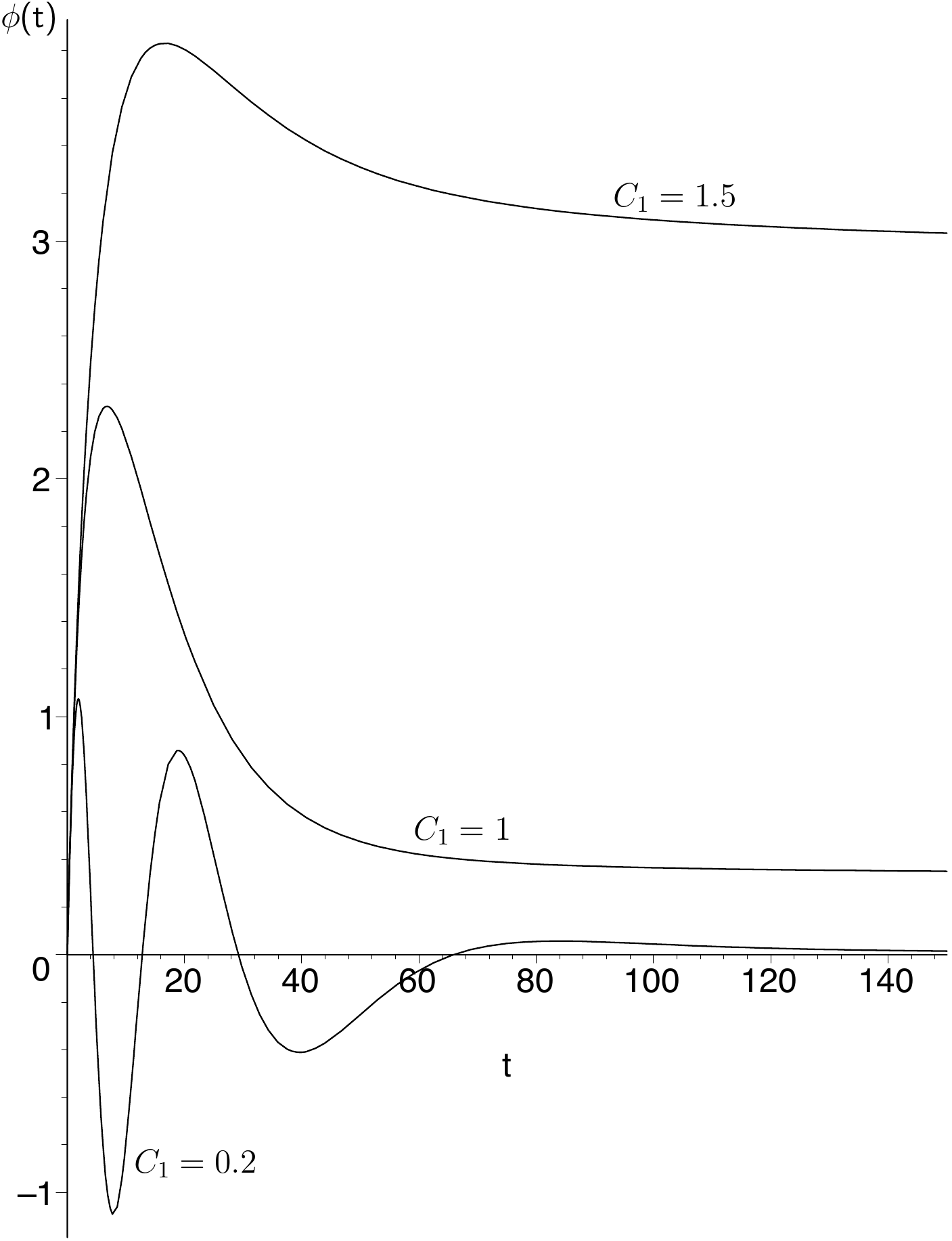,width=.45\textwidth}
\caption{Influence of the stretch rate $C_1$ in the 
$X$-direction on the shearing motion solutions for a 
Blatz-Ko material (foamed, polyurethane elastomer).}
\end{figure}

Now, consider the \textit{generalized sinusoidal standing waves} of Section
III.B. As seen there, they are governed by nonlinear ordinary differential
equations when the longitudinal displacement $u$ is such that $u_X = U =
U(X) $, a function of $X$ only. This condition leads to \eqref{cm0b}, here
integrated twice as
\begin{equation}
(\phi ^{^{\prime }2}+\psi ^{^{\prime }2}-1)U^{-3}+U^{\frac{(\beta -2)}{\beta
}}=C_{3}X+C_{4},  \label{h14}
\end{equation}
where $C_{3}$, $C_{4}$ are disposable constants. Taking the typical value $%
\beta =2$ corresponding to Poisson ratio $\nu =1/4$, we find
\begin{equation}
U=\left[ \dfrac{\phi ^{^{\prime }2}+\psi ^{^{\prime }2}-1}{C_{3}X+C_{4}-1}%
\right] ^{\textstyle{\frac{1}{3}}}.  \label{h15}
\end{equation}
Hence for this material, the governing equations \eqref{cm0} for the
transverse solutions of type \eqref{ses14b} at $\theta (X)=0$ are, using %
\eqref{h15} and \eqref{h7a}$_{1}$,
\begin{align}
& \left[ \mu \left( \dfrac{\phi ^{^{\prime }2}+\psi ^{^{\prime }2}-1}{%
C_{3}X+C_{4}-1}\right) ^{-\textstyle{\frac{2}{3}}}\phi ^{\prime }\right]
^{\prime }+\rho _{0}\omega ^{2}\phi =-\eta \omega \psi ^{\prime \prime },
\notag \\
& \left[ \mu \left( \dfrac{\phi ^{^{\prime }2}+\psi ^{^{\prime }2}-1}{%
C_{3}X+C_{4}-1}\right) ^{-\textstyle{\frac{2}{3}}}\psi ^{\prime }\right]
^{\prime }+\rho _{0}\omega ^{2}\psi =\eta \omega \phi ^{\prime \prime }.
\end{align}
The final governing equations are highly nonlinear, even with the choice $%
\beta =2$. The possibility of solving them in closed form seems quite
remote. Of course, numerical and qualitative analyses may be performed with
the usual methods of dynamical systems theory.


\subsection{A separable strain energy density}


Recall that the strain energy function of an \textit{incompressible}
material, $\Sigma_\text{inc}$ say, corresponds to the restriction to the
subspace $(I_1, I_2, 1)$ of a strain-energy $\widetilde{\Sigma}$ say,
defined in the full space $(I_1, I_2, I_3)$:
\begin{equation}
\Sigma_\text{inc} (I_1, I_2):= \widetilde{\Sigma}(I_1, I_2,1).
\label{restrict}
\end{equation}
Hence the classical incompressible neo-Hookean form,
\begin{equation}
\Sigma_\text{inc} = \mu(I_1 - 3)/2,  \label{neo}
\end{equation}
may be associated for example with the full-space strain energy
\begin{equation}
\widetilde{\Sigma} = \frac{\mu}{2} (I_1 - 3 - 2\ln J), \quad J:= \sqrt{I_3},
\end{equation}
often used in the classical molecular theory of rubber \cite{Erman1}.

Now, constitutive equations for \textit{compressible} hyperelastic materials
come in many formulations. One of them consists in adding a purely
volumetric term $\Sigma_\text{vol}(J)$ say, to a basic strain energy density
function $\widetilde{\Sigma}$, whose restriction \eqref{restrict} is the
strain energy function of an incompressible material $\Sigma_\text{inc}$.
Then the final strain energy density for a compressible hyperelastic
material can be written as
\begin{equation}
\Sigma = \widetilde{\Sigma}(I_1,I_2,I_3) + \Sigma_\text{vol}(J).  \label{se2}
\end{equation}

Several models have have been proposed in the literature for this \emph{pure
volumetric part} (or \emph{bulk term}) of the strain energy function. Ogden
\cite{O73} proposed the form:
\begin{equation}
\Sigma_\text{vol}^I (J) = \lambda \beta^{-2}(\beta \ln J + J^{-\beta} - 1),
\label{st5}
\end{equation}
where $\lambda$ is the first Lam\'e modulus and $\beta$ ($>0$) is an
empirical parameter. Flory \cite{flory} proposed the form
\begin{equation}
\Sigma_\text{vol}^{II}(J) = (c/2)(\ln J)^2,
\end{equation}
and Simo and Pister \cite{simopis}, the form
\begin{equation}
\Sigma_\text{vol}^{III}(J)= (c/2)[(J^2 - 1) - \ln J],
\end{equation}
which is \eqref{st5} at $\beta = -2$. Recently Bischoff et al. \cite
{Bischoff} proposed the form
\begin{equation}
\Sigma_\text{vol}^{IV}(J)= \frac{c}{\beta^2}[\cosh \beta(J-1) - 1].
\end{equation}

In this manner, several strain energy functions may be constructed. As an
example, consider the power-law function of Knowles \cite{Kn},
\begin{equation}  \label{power}
\Sigma_\text{inc} = \dfrac{C}{b} \left[\left(1 + \dfrac{b}{n}(I_1 -
3)\right)^n - 1 \right],
\end{equation}
where $C$, $b$, and $n$ are constitutive parameters, all three assumed
positive. This simple strain energy captures several important features of
rubber-like materials. At $n=1$, it recovers the neo-Hookean form \eqref{neo}%
; at $n>1$ it describes strain-stiffening materials; at $n<1$,
strain-softening materials. These hyperelastic properties prove crucial to
the understanding of complex phenomena such as dynamic fracture, as shown
recently by Buehler et al \cite{BAG}. Then, using the formulation exposed
above, we may construct the following compressible strain energy function
\begin{equation}  \label{ses2b}
\Sigma = \frac{C}{b} \left[ \left( 1+\frac{b}{n}(I_{1}-3)\right)^n - 1 %
\right] -C\ln J+\Sigma _{\text{vol}}(J).
\end{equation}

Turning back to the finite amplitude motions of Section II, we find that $J =
\sqrt{I_3} = u_X = U$, so that $\Sigma_\text{vol}(J) = \Sigma_\text{vol}(U)$%
, and the functions $Q_1$, $Q_2$ defined in \eqref{Q1Q2} are here,
\begin{align}
& Q_1 = 2C [1 + \dfrac{b}{n}(U^2 + \Omega^2 - 1)]^{n-1}, \notag \\
& Q_2 = U^{-1} \Sigma_\text{vol}^{\prime}(U) - C U^{-2}.
\end{align}
In general, the corresponding equations of motion are coupled and rather
involved. Consider the special case of the standing waves \eqref{ses23},
with the choice $U(t) = C_2$ ($C_1 = 0$ in \eqref{ses15}). Then $\phi(t)$
satisfies \eqref{ses24}, which is here
\begin{equation}
\phi^{\prime\prime}+ \frac{k^2 \eta}{\rho_0} \phi^{\prime}+ 2\frac{k^2 C}{%
\rho_0}[1 + \dfrac{b}{n}(C_2^2 - 1 + k^2 \phi^2)]^{n-1}\phi = 0.
\label{power2}
\end{equation}

We notice that this equation may admit not only the static solution: 
$\phi := 0$, but also the nontrivial static solutions:
\begin{equation}  \label{static}
\phi = \pm (1/k)\sqrt{1 -C_2^2 - n/b}.
\end{equation}
These solutions are real if and only if $b(1-C_{2}^{2})-n>0$: when the body is
compressed in the $X$-direction ($u = C_2X$, $C_2 < 1$), then the solution
is real if $b > n/(1 - C_2^2) > 0$; when the body is stretched ($u = C_2 X$,
$C_2 > 1$), the solution is not real (recall that Knowles \cite{Kn} assumed
that $b>0$ in \eqref{power}).

Another important general property of equation \eqref{power2} is that it can
be derived \cite{Steeb} from Lagrangian
\begin{equation}
\mathcal{L} = \textstyle{\frac{1}{2}} e^{\frac{k^2 \eta}{\rho_0} t}
[\phi^{\prime 2} - \mathcal{V}(\phi)],  \label{l1}
\end{equation}
where
\begin{align}  \label{l2}
& \mathcal{V}(\phi) = 2\dfrac{k^2 C}{\rho_0} \int_{0}^\phi[1 + \dfrac{b}{n}%
(C_2^2 - 1 + k^2 \zeta^2)]^{n-1} \zeta \text{d}\zeta \notag \\
& \quad \quad =\frac{C}{\rho_0 b}[1 + \dfrac{b}{n}(C_2^2 - 1 + k^2
\phi^2)]^n.
\end{align}

As an example we consider a \textit{strain-stiffening material}, by taking $%
n=2$ in \eqref{ses2b}. Then \eqref{power2} is a Duffing equation with
damping,
\begin{equation}
\phi^{\prime\prime}+ \delta \phi^{\prime}- \beta \phi + \alpha \phi^3 =0,
\label{se12}
\end{equation}
where
\begin{equation}  \label{se13}
\delta := \dfrac{k^2 \eta}{\rho_0}, \quad \beta:= \dfrac{k^2 C}{\rho_0} [b(1
- C_2^2) - 2], \quad \alpha:= b \dfrac{k^4 C}{\rho_0}.
\end{equation}
Clearly, $\alpha>0$, $\delta>0$. To make the connection with results by
Holmes \cite{Holmes}, we impose $\beta > 0$ also, which happens only when
the body is \textit{compressed} 
in the $X$-direction ($u = C_2X$, $C_2<1$) and when 
$b > 2/(1 - C_2^2)$. 
As seen for \eqref{se12}, this equation possesses three
fixed points in the phase plane, namely: $(0, 0)$, which is a saddle, and $%
(\pm \sqrt{\beta / \alpha}, 0)$, which are two sinks. Holmes \cite{Holmes}
showed that equation \eqref{se13} is locally and globally stable. It follows
that as $t\rightarrow \infty$, our standing waves must approach one of the
four wrinkled configurations
\begin{align}
& \lim_{t\rightarrow \infty}v(X,t) = \pm
\sqrt{\beta/\alpha}\cos(kX - \theta_0), \notag \\
& \lim_{t\rightarrow \infty}w(X,t) = \pm \sqrt{\beta / \alpha}
\sin(kX - \theta_0), \label{se13b}
\end{align}
for almost any initial conditions. As any one of the four final
configurations is equivalent to another, we conclude that the material is
destabilized by the waves (only some special choices of the initial
conditions will lead to the unstressed deformation corresponding to the
saddle point (0,0); these special initial conditions are found by computing
the corresponding stable manifold in the phase plane, and we refer to Holmes
\cite{Holmes} for the details.)

The situation described here can be extended to all positive integers $n$,
because the leading of nonlinearity of the potential $\mathcal{V}(\phi)$ in %
\eqref{l2} is $\phi^{2n}$ so that the leading order of nonlinearity in the
resulting differential equation is always to an odd integer power. We also
point out that for a given $b$, an increasing $n$ requires a decreasing $C_2$
for the appearance of the fixed points \eqref{static} that is, a greater
compressive stretch in the $X$-direction. This is in agreement with the
``physical'' expectation that the stiffer a material is, the harder it is to
destabilize it.

For non-integer powers $n$, the situation is more complex because
destabilizing configurations may or may not appear in the case of
strain-hardening materials ($n>1$) \textit{and} in the case of
strain-softening materials ($n<1$). Obviously the dynamical systems
uncovered here are susceptible to a more complete and more careful analysis than
the one provided, which may reveal even richer behaviors than those already
described, such as for example the appearance of parametric resonance; these
aspects are however beyond the scope of the present paper.

Finally, we note that the instability evoked in this Subsection is 
somewhat reminiscent of the buckling of an elasticita under a compressive 
load, but we point out several differences. 
The buckling of an elastica is a phenomenon related to geometrical 
non-linearities, whereas our instability is due to constitutive 
non-linearities;
in an elastica, more than one inflexion may occur whereas here, 
we could multiply instabilities only by considering a multiple well 
strain energy function 
(and such a strain energy is not usual for the modeling 
 of soft tissues and elastomers); 
for an elastica, the stability of the various fixed points must 
be examined thoroughly, whereas here, the stability of such points 
comes out directly from our dynamical approach.


\subsection{Fourth-order elasticity}


In the linearized theory of ``small-but-finite'' amplitude waves, the strain
energy function must be expanded up to the fourth order in the strain, in
order to reveal nonlinear shear waves \cite{Zabo86}. In that framework,
Murnaghan's expansion \cite{Murn51} is often used (see for instance Porubov
\cite{Poru03}):
\begin{equation}  \label{4thOrder}
\Sigma = \dfrac{\lambda + 2\mu}{2} i_1^2 - 2\mu i_2 + \dfrac{l+2m}{3} i_1^3
- 2m i_1i_2 + n i_3 + \nu_1 i_1^4 + \nu_2 i_1^2i_2 + \nu_3 i_1 i_3 + \nu_4
i_2^2,
\end{equation}
where $\lambda$, $\mu$ are the Lam\'e moduli, $l$, $m$, $n$ are the
third-order moduli, and $\nu_1$, $\nu_2$, $\nu_3$, $\nu_4$ are the
fourth-order moduli (other notations exist: see Norris \cite{Norris} for the
connections). In this expansion, we used the first three principal
invariants $i_1$, $i_2$, $i_3$ of $\mathbf{E}$, the Green-Lagrange strain
tensor; they are related to the first three principal invariants $I_1$, $I_2$%
, $I_3$ of $\mathbf{B}$ defined in \eqref{invariants} by the relations \cite
{Beat87},
\begin{align}  \label{invariants4}
& I_1 = 2i_1 + 3, \notag \\
& I_2 = 4i_1 + 4i_2 + 3, \notag \\
 & I_3 = 2i_1 + 4i_2 + 8i_3 + 1.
\end{align}

By chain rule differentiation, we find that the functions $Q_1$ and $Q_2$
defined in \eqref{Q1Q2} are here,
\begin{align}  \label{coeff1}
& Q_1 = \mu + (\lambda + 2\mu + m) i_1 + (l + 2m - \dfrac{\nu_2}{2})i_1^2
\notag \\
& \quad \quad \quad \quad - (2m + \nu_4)i_2 + 4 \nu_1 i_1^3 + 2\nu_2 i_1 i_2
+ \nu_3 i_3,  \notag \\
& Q_2 = -\mu - m i_1 + \dfrac{\nu_2}{2} i_1^2 + \nu_4 i_2,
\end{align}
where now $i_1$, $i_2$, $i_3$ are found from \eqref{invariants4} and %
\eqref{invariants3} as
\begin{equation}  \label{i_i}
i_1 = (U^2 + \Omega^2 - 1)/2, \quad i_2 = -\Omega^2/4, \quad i_3 = 0.
\end{equation}

Several comments are in order at this point. First of all, we recall that
the equations corresponding to (\ref{equazio}) for third-order elasticity
(without dissipation) were first derived by Gold'berg \cite{Gold} in 1960.

Then we note that recently there has been a renewed interested in
fourth-order elasticity following experiments on the nonlinear acoustic
properties of soft tissue-like solids \cite{CGF}. Hence Hamilton et al. \cite
{HZ} proposed the following reduced version of the expansion \eqref{4thOrder}%
, suitable when the relative portion of energy stored in compression is
negligible, as is expected for shear deformations and motions of soft
tissues,
\begin{equation}  \label{hamil}
\Sigma = -2 \mu i_2 + n i_3 + \nu_4 i_2^2.
\end{equation}
In this case, the functions $Q_1$ and $Q_2$ defined in \eqref{Q1Q2} simplify
to
\begin{equation}
Q_1 = - Q_2 = \mu - \nu_4 i_2 = \mu + \nu_4 \Omega^2/4,  \label{coeff2}
\end{equation}
and we check at once that the longitudinal equation of motion \eqref{equazio}%
$_1$ becomes a trivial identity in the purely elastic case ($\zeta = \chi =
\eta = 0$), because here $Q_1+Q_2=0$.

Finally, we also notice that the theories of third- and fourth-order
elasticity have been used to study bulk solitons propagation in elastic
materials. For instance, Hao and Marris \cite{Hao} discuss the issue of
acoustic solitons from both experimental and theoretical points of view;
they consider the possibility to produce KdV solitons by adding a
fourth-order spatial derivative to the longitudinal wave equation and by
performing some \emph{ad hoc} approximations.

The general expressions \eqref{4thOrder}, \eqref{coeff1}, and \eqref{i_i}
yield quite complicated equations of motion. If for example we restrict our
attention to the standing waves \eqref{ses23}, we find that the polynomial
nonlinearity of the resulting ordinary differential equation is of the fifth
order for third-order elasticity, and of seventh order for fourth-order
elasticity. If we further restrict our attention to the reduced expansion %
\eqref{hamil} we find that equation \eqref{ses24} reduces to a damped
Duffing equation,
\begin{equation}
\phi ^{\prime \prime }+\dfrac{\eta k^{2}}{\rho _{0}}\phi ^{\prime }+\dfrac{%
k^{2}}{4\rho _{0}}(4\mu +k^{2}\nu _{4}\phi ^{2})\phi =0.  \label{df}
\end{equation}
This equation is characterized by a positive linear stiffness and therefore
the peculiar behavior found in the previous sub-section for some power-law
materials is \emph{ruled out}. For completeness, we point out that Holmes and Rand
\cite{HR} computed an approximate solution of equation \eqref{df}, using the
method of averaging.


\section{Concluding Remarks}


We studied the propagation of finite amplitude waves in a general nonlinear
elastic material, with dissipation taken into account by means of a simple
mechanism of differential type. We showed that the general equations of
motion admit some beautiful and most interesting reductions to ordinary
differential equations, using a direct method based on complex functions.
Clearly, these reductions are a consequence of the inherent symmetrical
structure of the balance and constitutive equations and may therefore be
recovered also by the standard methods of group analysis; we argue however
that our direct method is more simple and more revealing from a mechanical
point of view. Moreover, our results complement and generalize the
well-known results of Carroll in various directions, as already pointed out
in the Introduction; they also emphasize the analogy between Carroll's
solutions and some exact solutions proposed for nonlinear strings by Rosenau
and Rubin \cite{RR}.

The solutions provided here can be used not only as benchmarks for the more
complicated numerical analysis required in ``real life'' applications, but
also for a better understanding of the mechanical properties and of the
mathematical structure of various usual models. These advantages were
highlighted in the discussion on standing waves solutions conducted in
Section III. There, we focused much of the discussion on the special case of
standing waves for the sake of simplicity and brevity, but it is clear that
our general methods of investigation can be applied to all the ordinary
differential equations that were derived.

One of the important findings uncovered by the analysis of our solutions is
the possibility to predict the appearance of highly symmetric wrinkles in an
elastic medium compressed longitudinally. This phenomenon has been observed
experimentally by several researchers when, for example, a hard film is
deposited on a soft material and put under compressive strain \cite{HHS}.
Our analysis showed that these wrinkles may indeed appear when the
compression parameter satisfies some simple inequalities; it also showed
that the detailed viscoelastic behavior of the material is unimportant
because of the highly attracting nature of the wrinkled states of
deformations. We note of course that our solutions have a higher symmetry
than the patterns usually observed in experiments, probably because our
solutions are characterized by strong invariance and because they are bulk
solutions, in contrast to the half-space geometry generally used in
experiments. Despite these limitations, we found that these solutions are
compatible with the relatively simple nonlinear power-law strain energy
density whereas weak non-linear theories (up to the fourth-order) are not
adequate to predict these features.

Our approach allowed a direct and complete comparison of the models used in
nonlinear acoustics and in continuum mechanics. It was seen that although
the popular weak nonlinear theories are often used to study some special
features of wave propagation, a consequence of their polynomial nature is
that they cannot contain all the necessary information of the general
constitutive theory.

Last but not least, we gave some explicit solutions to some models of
compressible hyperelasticity. New exact solutions are always welcomed and
valuable additions to the short list of exact solutions found in the
literature.

We conclude by pointing out that several generalizations of our results are
possible in principle. For instance, the consideration in polar coordinates
of a wave propagating in the radial direction coupled to two shear waves in
the axial and azimuthal direction is one of such possible generalizations.
In this case the following class of motions is considered
\begin{equation}
r=r(R,t),\quad \theta =\Theta +g(R,t),\quad z=Z+u(R,t),  \label{p1}
\end{equation}
where $(R,\Theta ,Z)$ are cylindrical polar coordinates associated
with the undeformed configuration and $(r,\theta ,z)$ are
cylindrical polar coordinates associated with the deformed
configuration. It is possible to show that from \eqref{p1} and the
corresponding kinematical quantities of interest, the equations of
motion may be reduced to a set of three partial differential
equations, similar to what we did for the motions \eqref{m1} in
\eqref{equazio}. Unfortunately, it seems that in this case the
equations of motion do not allow general solutions in separable
form, as was the case in Section III for the equations
\eqref{equazio}). Of course, we do not exclude that for special
classes of materials, it is possible to find some classes of wave
solutions (see for example \cite {HT} for compressible materials
and \cite{FO} for incompressible materials).


\end{document}